\documentclass{article}
\usepackage[utf8]{inputenc}
\usepackage[T1]{fontenc}
\usepackage[english]{babel}
\usepackage{graphicx}
\usepackage{floatflt}
\usepackage{amsthm}
\usepackage{amsmath}
\usepackage{amssymb} 
\usepackage[usenames, dvipsnames]{color}
\usepackage{mathrsfs}
\usepackage{geometry}  
\usepackage{makeidx}
\usepackage{physics}
\usepackage{mathtools}
\usepackage{authblk}
\usepackage{lineno}
\usepackage[numbers,sort&compress]{natbib}
\usepackage{csquotes}
\usepackage{chngcntr}
\usepackage{setspace}
\usepackage[labelfont=bf]{caption}
\usepackage{comment}
\usepackage{ulem}

\DeclareUnicodeCharacter{03BC}{$\mu$}
\DeclareUnicodeCharacter{2009}{ }
\DeclareUnicodeCharacter{03BB}{\ensuremath{\lambda}}
\DeclareUnicodeCharacter{223C}{\ensuremath{\sim}}
\DeclareUnicodeCharacter{2215}{}

\usepackage{xr-hyper}
\usepackage[hidelinks]{hyperref}
\makeatletter
\newcommand*{\addFileDependency}[1]{
  \typeout{(#1)}
  \@addtofilelist{#1}
  \IfFileExists{#1}{}{\typeout{No file #1.}}
}
\makeatother
\newcommand*{\myexternaldocument}[1]{%
    \externaldocument{#1}%
    \addFileDependency{#1.tex}%
    \addFileDependency{#1.aux}%
}
\myexternaldocument{Supplementary}
\usepackage{booktabs}
\usepackage{makecell}
\usepackage{multirow}

\usepackage[table]{xcolor}

\definecolor{lightgray}{gray}{0.9}

\usepackage{titlesec}

\renewcommand{\thesection}{\Roman{section}}
\renewcommand{\thesubsection}{\Alph{subsection}}
\renewcommand{\thesubsubsection}{\arabic{subsubsection}}

\titleformat{\section}
  {\normalfont\bfseries\MakeUppercase}{\thesection.}{1em}{}

\titleformat{\subsection}
  {\normalfont\bfseries}{\thesubsection.}{1em}{}

\titleformat{\subsubsection}
  {\normalfont\bfseries\itshape}{\thesubsubsection.}{1em}{}

\title{Epitaxial growth optimization, measurement and theoretical analysis of strain-compensated QCL grown on (511)A InP}

\author[1,*]{A. Cargioli}
\author[1]{M. Beck} 
\author[1]{J. Faist}

\affil[1]{Institute for Quantum Electronics, ETH Zurich, CH-8093 Zurich, Switzerland}

\affil[*]{Contact Email: acargioli@phys.ethz.ch}

\date{}

\modulolinenumbers[1]

\begin{document}
\renewcommand{\figurename}{Figure}
\def\equationautorefname#1#2\null{Eq.#1(#2\null)}
\maketitle
\numberwithin{equation}{section}

\begin{abstract}
Interface roughness scattering is an important limiting factor for achieving high performance Quantum Cascade Lasers. Following recent results, we study the growth conditions for a strain-compensated QCL emitting around 4.6 $\mu$m grown on a (511)A InP substrate using AFM and XRD measurements. We find that modulating the arsenic flux and correctly tailoring the III/V ratio is fundamental to achieve a good quality material. We report the first lasing device on such a platform with a current density threshold of 1.34 kA/cm$^2$ and a slope efficiency of 1.1 W/A, which result suboptimal compared to the (100) reference. Finally, we find a 7\% redshift of the (511)A spectrum which we attribute to an impurity scattering due to the increased incorporation along the exposed (111) direction. We validate this statement by verifying that the change in CBO and effective electron mass due to strain along a non-trivial direction cannot cause such a shift by using the $\textbf{k}\cdot\textbf{p}$ method generalized to arbitrary growth directions. 

\end{abstract}

\counterwithout{equation}{section}

\newpage

\section{Introduction}

\begin{figure}[htb!]
    \centering
    \includegraphics[width=\linewidth]{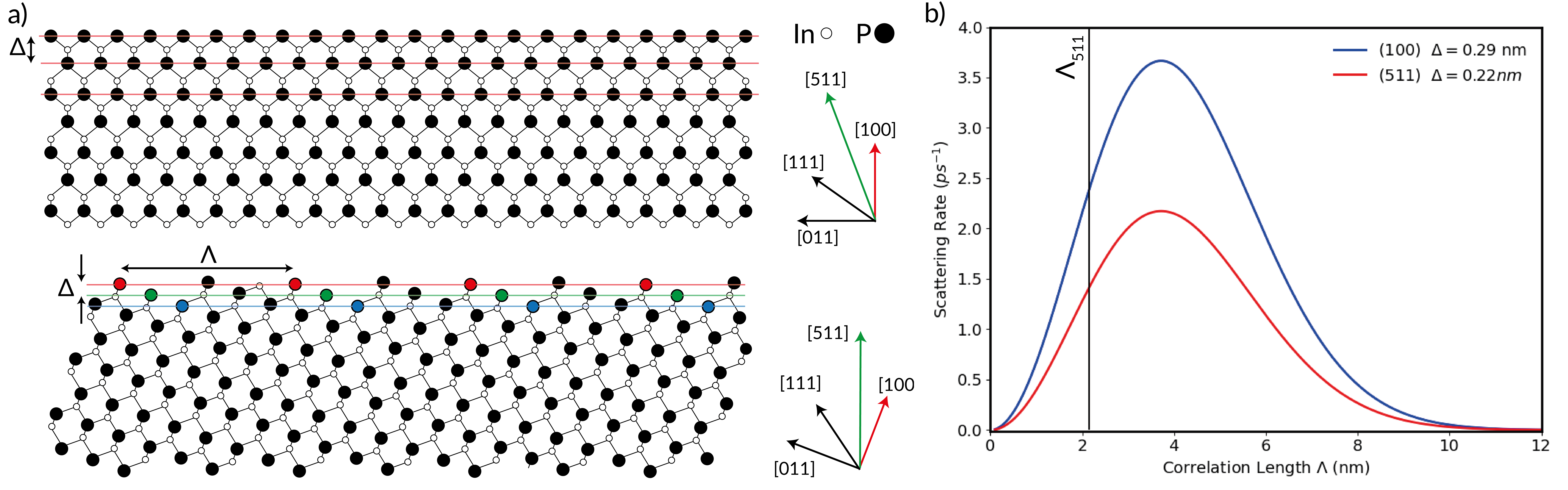}
    \caption{a) Schematic of the InP lattice oriented along the (100) and (511) direction. In both cases the monolayer planes are highlighted. b) Computed scattering rate for a transition energy of 270meV for the (100) and (511) case, as a function of the correlation length and assuming as $\Delta$ the monolayer thickness.}
    \label{fig:lattice}
\end{figure}

Among the different mechanisms influencing the performance of Quantum Cascade Lasers (QCLs), interface roughness is one of the most relevant contributing to the laser transition broadening \cite{khurgin_role_nodate, khurgin_inhomogeneous_2008, franckie_impact_2015,flores_leakage_2013, deutsch_probing_2013} and the intersubband lifetime. This contribution becomes particularly significant for laser transitions exceeding 250 meV ($\lambda<5\mu$m), even at room temperature. A systematic study, which correlated the role of interface roughness with the laser performance was performed by Bismuto \textit{et al.} \cite{bismuto_influence_2011} on  devices emitting at 4.6 $\mu$m by changing the epitaxial growth temperature. In general, it is important to accurately control the growth condition to ensure good material quality and reduce surface roughness scattering. Generally, the models describing the interface roughness scattering assume the statistical nature of the surface morphology with a gaussian or exponential correlation profile \cite{unuma_intersubband_2003} described by the height step parameter $\Delta$ and the correlation length $\Lambda$. Typical values reported in the literature are of the order of a crystal monolayer for $\Delta$ (few \AA) and of the order of 4-12 nm for $\Lambda$ \cite{tsujino_interface-roughness-induced_2005, bismuto_influence_2011}, although most of the time the estimation relates to the value of the product $\Delta\Lambda$, since decoupling the two parameters is non-trivial.  Among the different techniques used to reduce the interface roughness, such as high-temperature growth or III/V ratio optimization, there is the possibility of using different crystal orientations compared to the standard (100). In this case, the substrate surface will not be terminated by a single atomic monolayer, but will be composed by alternating "terraces" \cite{hiyamizu_extremely_1994} where their height and distance are fixed by the material and the crystal orientation, and not by the random rearrangement of the atoms at the interface. In Fig.\ref{fig:lattice}.a we report a simple schematic showing the InP crystal structure, highlighting the atomic monolayer planes of a (100) and (511) oriented crystal. We assume that the typical height jump between monolayers will be half of the unit-cell length in the (100) case ($\Delta = 0.29$ nm) while, in the (511) case, it will be only a fraction of that ($\Delta = 0.22$ nm) due to the reduced angle. In this case, we notice that, in addition to the reduction of the monolayer distance, we also introduce a different periodicity on the surface which will fix $\Lambda = 2.15$ nm, contrary to what happens for (100) crystals, where the parameter depends on the growth condition. 

Previous attempts of heterostructures grown on non-trivial crystal orientations showed an improvement in the width of the electroluminescence peak \cite{hiyamizu_extremely_1994, shimomura_extremely_1995}, and in the electron mobilities \cite{masataka_higashiwaki_dc_2000}. More recently, Semtsiv \textit{et al.} \cite{semtsiv_reduced_2018} demonstrated improved performance for a lattice-matched InGaAs/AlInAs QCL structure grown on (411)A InP emitting at 9 $\mu$m. Specifically, they reported a 2 to 3 time increase in slope efficiency, a lower threshold current, and growth conditions that were comparable to those used for the (100) conventional orientation. It is therefore important to understand if it is possible to replicate such results for shorter wavelengths. In particular, we can look at the interface roughness scattering rate for an energy close to the laser transition, as a function of the correlation length in Fig.\ref{fig:lattice}.b for the two crystal directions. In both cases, all parameters are the same except for the different monolayer distance (see also Supplementary Information \ref{supp:scattering}). Since it is not easy to estimate a priori the value of $\Lambda$ for the (100) growth, it is therefore worth investigating whether growing in a direction where both parameters are chosen by the crystal orientation brings an improvement to the laser performance.  In this work, we first report a study on the growth conditions required to grow strain compensated QCLs on (511)A oriented InP substrate. The (411)A direction was initially explored by it was not possible to grow high quality heterostructures because of the difficulty of finding a good growth parameter window. Subsequently, we report results on the lasers grown on the (511)A orientation which are directly compared with control lasers with the same nominal active material grown on a conventional (100) InP substrate. Finally, we analyze the spectral emission which presents a considerable red-shift compared to the emission of the control lasers. Since we are growing strained material on a non-trivial direction, it is not obvious whether the conduction band offset (CBO) or the effective electron mass would present a significant change compared to the standard direction. Therefore, we report a complete derivation of the CBO and effective mass for an arbitrary growth direction, using the results of the work of Yang \textit{et al.} \cite{yang_strain_1994} and Suguwara \textit{et al.}  \cite{sugawara_conduction-band_1993}, confirming that they cannot be the cause of the red-shift, which we ultimately attribute to impurity scattering \cite{faist_quantum_2013}.

\section{Optimization of the Growth Parameters}

\begin{figure}[htb!]
    \centering
    \includegraphics[width=.9\linewidth]{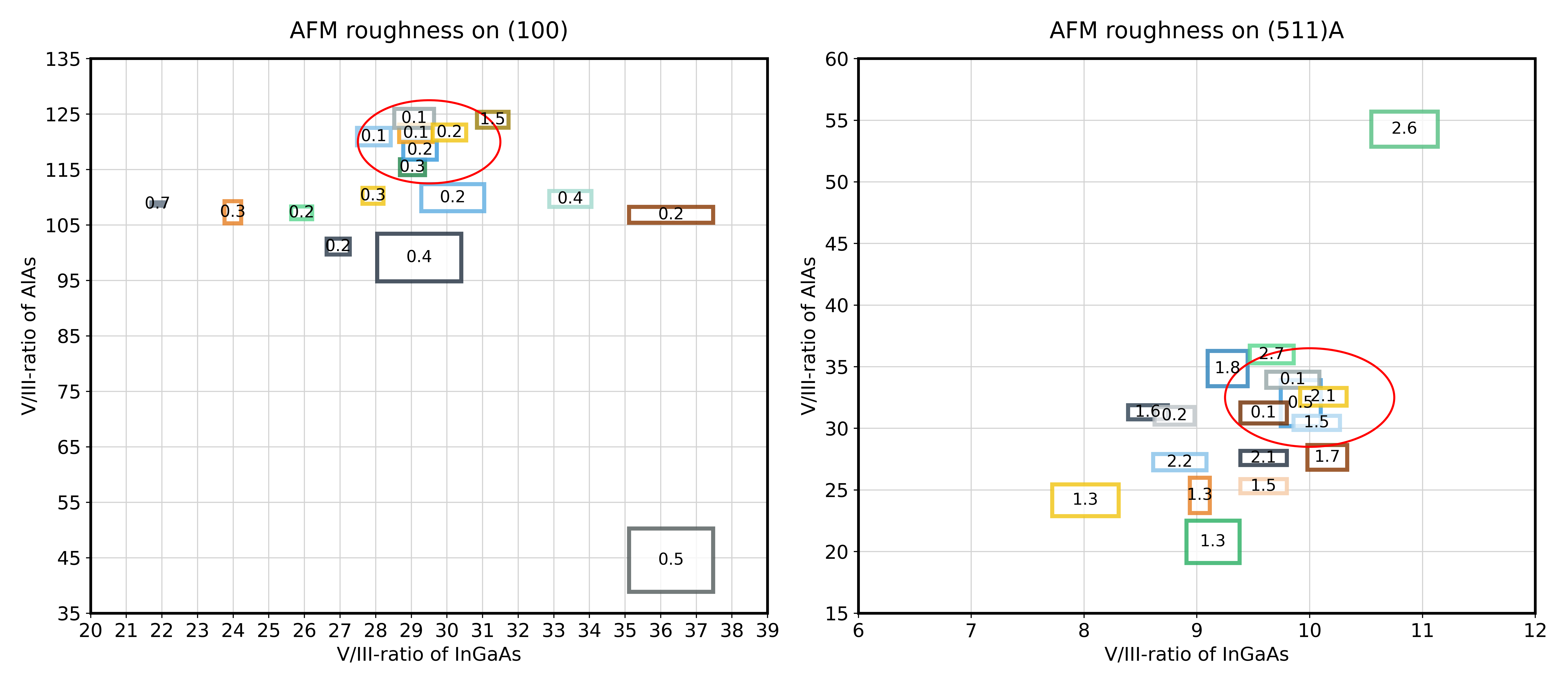}
    \caption{Correlation plots of surface roughness (nm) as a function of the V-III ratio of AlAs and InGaAs grown on (100) and (511) oriented InP substrates. The red line indicated the optimal growth windows.}
    \label{fig:53ratio}
\end{figure}

In contrast to the study conducted by Semtsiv \cite{semtsiv_reduced_2018}, the main difficulty is to determine whether high-quality strain compensated material can be grown along non-standard crystallographic orientations. The strain is required to increase the CBO and achieve shorter emission wavelengths. In particular, the structure used in this work (EV3032) is based on a pocked injector design and with a composition of In$_{0.674}$Ga$_{0.326}$As/Al$_{0.652}$In$_{0.348}$As and a nominal sheet carrier density of $1.25\cdot10^{11}$cm$^{-2}$ (more details are provided in Supplementary Information \ref{sec:active_region}). We first perform preliminary calibration growths of bulk InGaAs and AlInAs grown on (100), (511)A and (411)A orientations. We analyze the surface by Atomic Force Microscopy (AFM) measurements and, for lattice matched materials, we find a slight improvement for the (411)A samples, compared to the reference ones, but not for the (511)A direction (see Supplementary Information \ref{sec:afm}). Subsequently, we grow strain compensated calibration structures made of 75 periods of AlAs/In$_{0.7}$Ga$_{0.3}$As varying the V/III ratio for AlAs and In$_{0.7}$Ga$_{0.3}$As separately and assessing the best growth conditions using AFM and X-Ray Diffraction (XRD) measurements. In this case, it is not possible to find a growth window, making it impossible to achieve a good heterostructure quality for the (411)A direction. Therefore, the data reported are relative only to the (511)A direction and to the control (100) direction. In Fig.\ref{fig:53ratio} we report the surface roughness of various strain compensated samples as a function of the different V/III ratio for AlAs and InGaAs. The samples exhibiting the lowest surface roughness define the optimal range of growth parameters (red circle), which is observed to be narrower for the (511)A orientation.

\begin{figure}[htb!]
    \centering
    \includegraphics[width=0.6\linewidth]{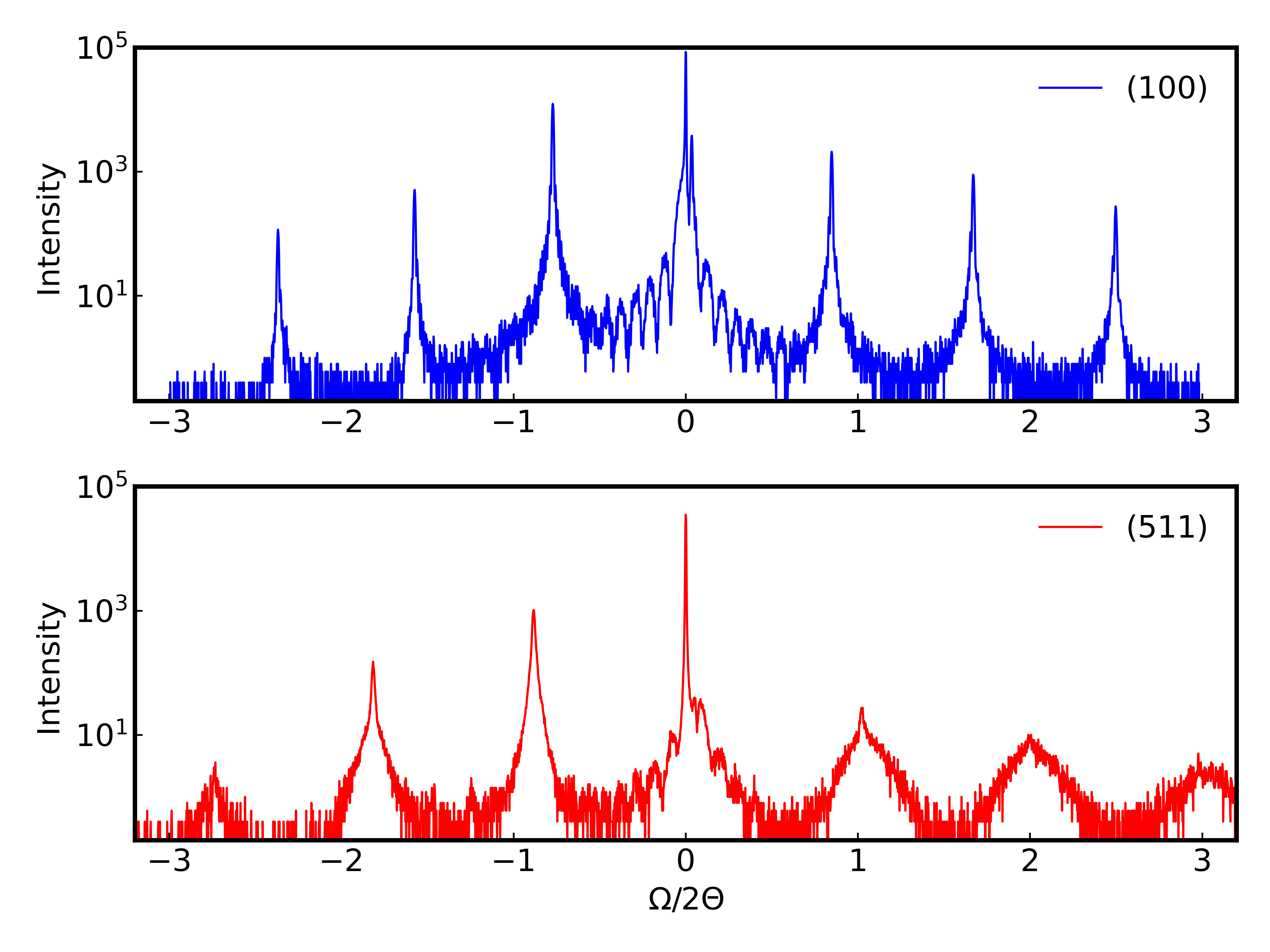}
    \caption{Example of an XRD measurement performed on a AlAs/In$_{0.7}$Ga$_{0.3}$As superlattice grown along a (100) and (511) direction.}
    \label{fig:xrd411}
\end{figure}

As a result, growing thick heterostructures along this direction becomes more difficult. In particular, due to the different arsenic incorporation, depending on whether we are growing a well or a barrier material, we have to modulate its flux. This requires a very accurate control of the shutters, which, combined with a narrower growth window, makes the growth of thick structures extremely challenging. We also directly compare a typical XRD measurement of a superlattice grown on a (100) substrate with the one grown on the (511)A direction, reported in Fig.\ref{fig:xrd411}.  In general, the diffraction peaks of the (511) sample seem to be less sharp, indicating a slightly worse structure periodicity, especially for positive angles, and possibly a less sharp material interfaces. Nevertheless, it is still possible to grow a full QCL laser structure on the (511)A oriented substrate and achieve lasing.

\section{Laser Characterization}
\subsection{Electrical Behavior}

\begin{figure}[b!]
    \centering
    \includegraphics[width=1\linewidth]{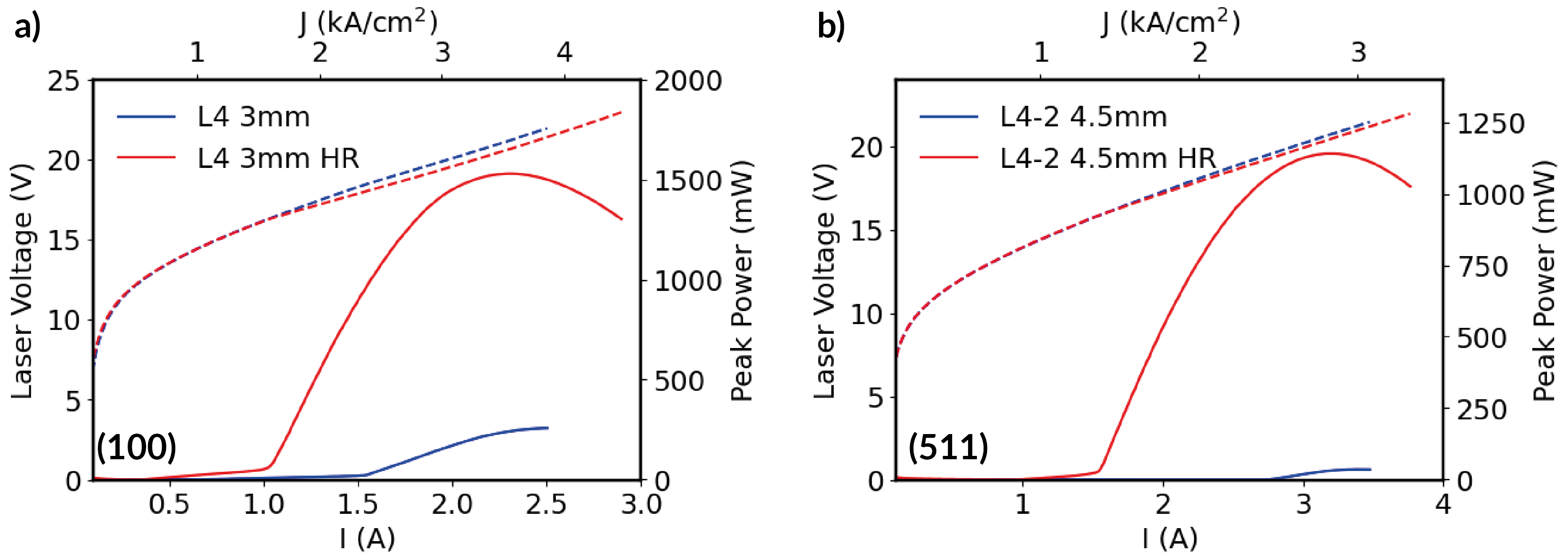}
    \caption{a) LIV curves of a laser grown on (100) before and after HR coating. b) LIV curves of a laser grown on (511) before and after HR coating. }
    \label{fig:liv511}
\end{figure}

The two layers under analysis are EV3032 and EV3033, nominally identical and grown on (100) and (511)A substrates, respectively. After growth, a Si-doped InP layer was grown by Metal Organic Vapor Phase Epitaxy (MOVPE) and a quarter wafer was processed in a ridge laser (see Supplementary Information \ref{sec:laser proc}). The lasers are evaluated in pulsed mode (2\% duty cycle) using a high-power MOSFET and at a heat-sink temperature of 253 K.  The Light-Current-Voltage (LIV) characteristics of the best-performing device for each growth are presented in Fig.\ref{fig:liv511}, both before and after the deposition of a high-reflection (HR) coating on the rear facet. The L4 laser, fabricated from the (100) sample, has dimensions of 3 mm $\times$ 21.5 $\mu$m, whereas the L4-2 laser, fabricated from the (511) sample, measures 4.5 mm $\times$ 25.1 $\mu$m. All relevant parameters extracted from the LIV measurements are summarized in Tab.\ref{tab:summarypars}. Considering first the data for the (511) laser, specifically the dynamic range before HR coating and the drastic increase in slope efficiency (from 77 to 1104 mW/A) after enhancing the mirror reflectivity, it becomes evident that the device operates in a “gain-starved” regime. More precisely, the effective doping level of the active layer is reduced by at least 20\%, as suggested by the comparison of the maximum current densities of the two samples.

\begin{table}[t!]
    \centering
    \resizebox{\textwidth}{!}{%
    \begin{tabular}{c|c|c|c|c|c|c|c|c}
         \makecell{Laser\\ } & \makecell{$J_{th}$ \\ (kA/cm$^2$) } & \makecell{$J_{max}$ \\ (kA/cm$^2$) } & \makecell{$dP/dI$ \\ (mW/A) } &\makecell{P.P. \\ (mW) } & \makecell{Wallplug \\ Eff. } & \makecell{$\eta_{tr}$ \\ } & \makecell{$\alpha_{wv}$ \\ (cm$^{-1}$) } & \makecell{$g'\Gamma$ \\ (cm/kA) }  \\
         \hline
         \makecell{L4 \\ (100)}  & 2.28  &3.84   &327     &257    &0.99$\%$   & 8 $\%$      &\makecell{ \\ \\ 0.92 } &  \makecell{ \\ \\ 2.75 }    \\
         \makecell{L4-HR\\(100)} & 1.53  &3.53   &1749    &1530   &7.6$\%$    & 26.3$\%$    &     &         \\
         \hline
         \makecell{L4-2 \\ (511)}  & 2.44  &2.99   &77      &36     &0.10$\%$   & 2 $\%$      &\makecell{ \\ \\ 0.66 } &  \makecell{ \\ \\ 1.25 }    \\
         \makecell{L4-2-HR\\(511)} & 1.34  &2.83   &1104    &1140   &4  $\%$    & 17$\%$      &     &         \\        
    \end{tabular}%
    }
    \caption{Summary of the most important parameters for two lasers grown on (100) and (511) before and after HR coating, respectively. All measurements are performed with a power MOSFET, with a 2$\%$ duty cycle at a heat-sink temperature of 253 K. We report, in the appearance order, current density threshold, maximum current density, slope efficiency, peal power, wallplug efficiency, quantum efficiency, waveguide losses and gain coefficient}
    \label{tab:summarypars}
\end{table}

As already noted, the two layers are nominally identical, so one would expect similar doping levels and thus comparable $J_{max}$, which is not observed. This discrepancy can be directly attributed to the way Si incorporates during growth along the (111) direction, where it more frequently substitutes group V instead of group III atoms, thus increasing the p-type background. In particular, for the (511)A surface, the step between adjacent terraces runs along the (111) direction (see also Fig.\ref{fig:lattice}.a), which is why this effect is especially relevant for this sample. Even taking this into account, the comparison still results unfavorable for the (511) sample, which systematically presents lower parameters and thus a worst performance.

\subsection{Spectral Behavior}

\begin{figure}[htb!]
    \centering
    \includegraphics[width=\linewidth]{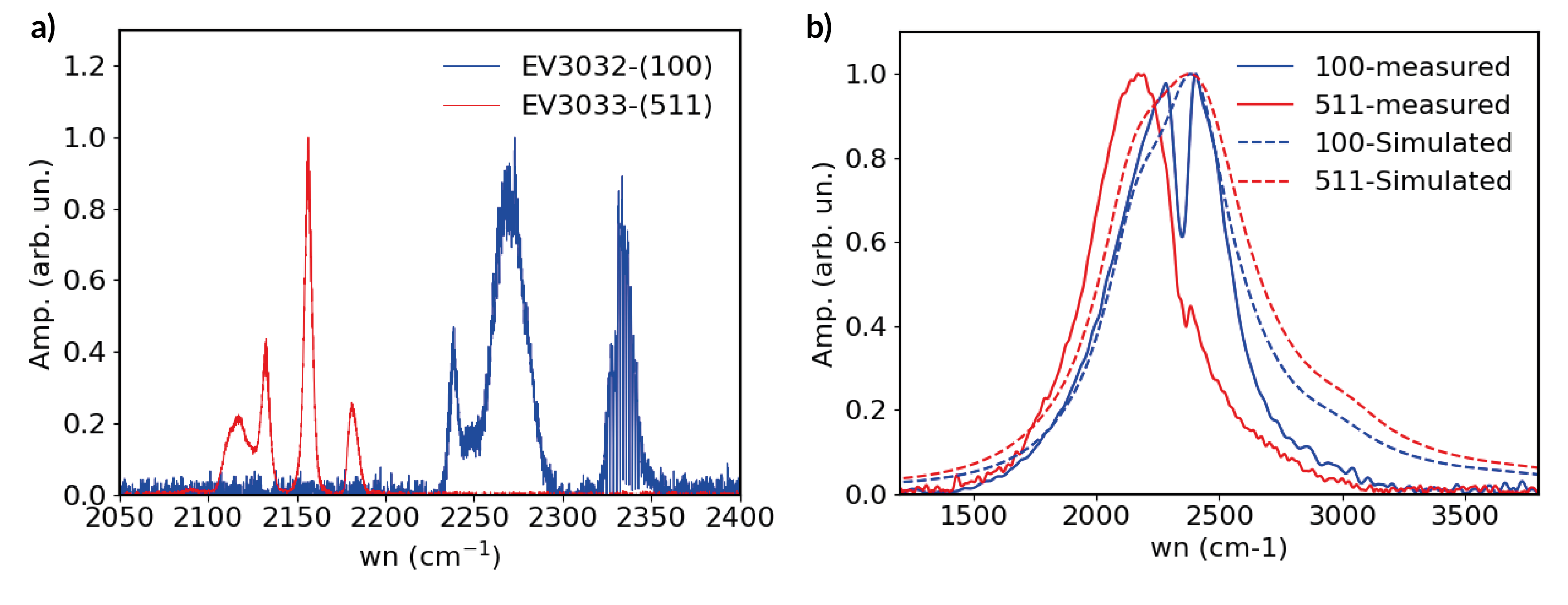}
    \caption{a) Laser spectra for the (100) and (511) sample acquired under pulsed operation at a heat-sink temperature of 253 K. b) Electroluminescence measured (solid lines) and simulated (dashed lines) for the (511) and (100) samples.}
    \label{fig:511spectra}
\end{figure}

We carry out spectral measurements in pulsed operation on the two lasers, as shown in Fig.\ref{fig:511spectra}.a, finding a pronounced red-shift for the (511) sample. The laser, designed to emit around 2300 cm$^{-1}$ ($\sim$ 285 meV), instead emits near 2150 cm$^{-1}$ ($\sim$ 265 meV), corresponding to an energy shift of about 7$\%$. In Fig.\ref{fig:511spectra}.b we also present subthreshold electroluminescence measurements on the same devices, to facilitate a more accurate comparison with the simulations (also shown in the figure) obtained with a density matrix formalism. Note that the dip in the center of the (100) luminescence peak does not originate from the device itself, but from the CO$_2$ absorption line, which could not be removed experimentally. Overall, the simulated luminescence for the (100) device agrees well with the experimental data, whereas the (511) simulation does not. This spectral shift cannot be ascribed to a change in active region thickness, which, according to XRD measurements, deviates from the expected value by $-0.63\%$ for the (100) sample and by only $-0.07\%$ for the (511). In addition, the bare rescaling of the full layer thickness would not explain the red-shift, and requiring a rescale of +7\% for the well thickness and -7\% for the barrier thickness is unrealistic given the well controlled growth conditions. Therefore, we attribute the change in emission wavelength to enhanced impurity-induced scattering, as also discussed in \cite{faist_quantum_1994}, chapter 7.9. Even if the two growth happened sequentially, it is still possible for the (511) sample to have incorporated more impurities compared to the (100) sample due to the exposed (111) plane, as already mentioned for Si doping. Although the latter is intentional doping, which happens only in some specific parts of the active region, other impurities in the MBE chamber, such as carbon and oxygen, can still be present, which, due to the inherent lower arsenic flux needed to grow good quality material, can substitute it. Given the particular nature of the growth direction and the fact that we are growing strained material, it is not obvious that the shift cannot also be attributed to the change in CBO or electron effective mass. Therefore, to exclude these factors and corroborate the statement regarding the impurities, we report the detailed calculation to compute the change in these parameters that are then used to perform the luminescence simulation reported in Fig.\ref{fig:511spectra}.b, which clearly does not match the measured one. This confirms that the effects induced by the strain for a growth along the (511)A are negligible.

\subsubsection{CBO Calculation}

The change in CBO is evaluated following the Van de Walle approach  \cite{van_de_walle_band_1989}, where we can compute the change in conduction and valence energy band due to strain following the relation

\begin{equation}
\label{eq:cbo2}
    \delta E_c = a_c \frac{\Delta \Omega}{\Omega} = a_c Tr(\hat{\epsilon}) \quad \text{where} \quad a_c = \frac{dE_c}{d\Omega}\Omega
\end{equation}

where $a_c$ is the "deformation potential" and $\hat{\epsilon}$ is the strain tensor that describes how the material deforms under the application of stress, in this case due to the lattice mismatch. The trace of the tensor evaluates the fractional volume change and can be fully computed using the elastic constants of the material. The total CBO change can be written as

\begin{equation}
\label{eq:cbo1}
    \Delta E_{CBO} = E_c^b + \delta E_c^b - E_c^w - \delta E_c^w
\end{equation}

where $E_c^{b,w}$ is the conduction band energy of the barrier and well material. Normally, the tabulated values in the literature refer to binary compounds, but since we are interested in ternaries as InGaAs or AlInAs we use the virtual crystal approximation to find the values for a given material composition, and then we correct for the strain using eq.\ref{eq:cbo2}-\ref{eq:cbo1}. TheCBO is fully determined by the deformation potential, which depends only on the material, and on the strain tensor which depends on the material and the growth direction. Therefore, we just need to write the tensor for a generic growth direction

\begin{equation}
    \mathbf{n} = \begin{pmatrix} n_1 \\ n_2 \\ n_3 \end{pmatrix} = \frac{1}{\sqrt{h^2 + k^2 + l^2}} \begin{pmatrix} h \\ k \\ l \end{pmatrix}.
\end{equation}

where the Miller index condition is used (\textit{hkl}). The full strain tensor is expressed as 

\begin{equation}
\label{eq:straintensor}
\epsilon_{ij} = \left( \delta_{ij} + n_i n_j \frac{A + \frac{1}{2} B (n_i^2 + n_j^2) + \frac{1}{2} C (n_i n_2 n_3)^2 (1/n_i^2 + 1/n_j^2)}{D + E (n_1^4 + n_2^4 + n_3^4) + F (n_1 n_2 n_3)^2} \right)\epsilon_\parallel 
\end{equation}

where $\epsilon_\parallel = (a-b)/b$, assuming $a,b$ the lattice constants for the substrate and the thin-film, respectively. The generic strain tensor is written following the result presented in the work of Yang \textit{et al.} \cite{yang_strain_1994}, and it is fully determined by the growth direction and by the different parameters from $A$ to $F$, which are functions of the elastic constants and are reported in the Supplementary Information \ref{sec:cbo}. For high symmetry directions, it is possible to find a relatively compact form of the tensor, which instead becomes quite cumbersome for arbitrary directions. In general, since it is symmetric, it is always possible to diagonalize it and find its eigenvalues and eigenvectors, which for the (511)A case can be written in the form

\begin{equation}
        \begin{pmatrix}
        1 & 0 & 0 \\
        0 & \lambda_1 -\lambda_2 & 0 \\
        0 & 0 & \lambda_1 + \lambda_2
\end{pmatrix}
\quad
v_1 = \begin{pmatrix} -1 \\1 \\0 \end{pmatrix}
\quad 
v_2 = \begin{pmatrix} \Lambda_1 + \Lambda_2 \\\Lambda_1 + \Lambda_2 \\1 \end{pmatrix}
\quad 
v_3 = \begin{pmatrix} \Lambda_1 - \Lambda_2 \\\Lambda_1 - \Lambda_2 \\1 \end{pmatrix}
\end{equation}

where $\lambda_{1,2}$ and $\Lambda_{1,2}$ are nontrivial functions of the elastic constants. For conciseness, the off-diagonal matrix elements and the complete forms of the eigenvalues and eigenvectors are provided in Supplementary Information \ref{sec:cbo}. The key point is not only that two eigenvectors are non-degenerate, which signals in-plane anisotropy, but also that the third eigenvector is not aligned with the growth direction. Instead, it forms an angle determined by the elastic constants and, therefore, by the material composition. This feature will be crucial for the effective mass estimation. Given the full tensor we can compute its trace and recover the change in CBO. In our case, for the EV3032 and EV3033 layer composition, we obtain a CBO of 827 meV for the (100) sample and 817 meV for the (511) sample. This corresponds to a relative variation of 1.2$\%$, which is insufficient to account for the 20 meV red-shift observed in the emission spectra in Fig.\ref{fig:511spectra}.

\subsubsection{Effective Mass Calculation}

The usual approach to compute the effective electron mass in the conduction band of a material is to use the $\textbf{k}\cdot\textbf{p}$ method, taking into account the interaction of conduction and valence states, as shown for the first time in \cite{kane_energy_1956,kane_band_1957}. We define the two conduction band states as $|S\uparrow\rangle$ and $|S\downarrow\rangle$, where the arrows indicate the spin state and $S$ the orbital part of the wavefunction. Instead, the valence states can be written in the form $|l,m\rangle$, where $l = 3/2$ or $l = 1/2$ for a total of 6 states which describe the Heavy-Hole, Light-Hole and Split-Off band. For a material without strain, the Kane Hamiltonian is built by writing the matrix elements using the $\textbf{k}\cdot\textbf{p}$  operator with the 8 possible states, which we report in detail in the Supplementary Information \ref{sec:effmass}. An important property of the operator is that it does not mix valence states between themselves but only with conduction states. In particular, all the non-trivial matrix elements can be expressed in 2 vectors as:

\begin{align}
    V_{states} &= \bigg( \left|\frac{3}{2}, \frac{3}{2}\right\rangle, \left|\frac{3}{2}, \frac{1}{2}\right\rangle, \left|\frac{3}{2}, -\frac{1}{2}\right\rangle, \left|\frac{3}{2}, -\frac{3}{2}\right\rangle, \left|\frac{1}{2}, \frac{1}{2}\right\rangle , \left|\frac{1}{2}, -\frac{1}{2}\right\rangle \bigg) \label{eq:vastates}\\
    \langle S\uparrow| \textbf{k}\cdot\textbf{p}| V_{states} \rangle &=  \bigg( Pk_+, -\sqrt{\frac{2}{3}}Pk_z, -\frac{1}{\sqrt{3}}Pk_-, 0, \frac{1}{\sqrt{3}}Pk_z, \sqrt{\frac{2}{3}}Pk_-\bigg) = s^+_j\\ 
    \langle S\downarrow| \textbf{k}\cdot\textbf{p}| V_{states} \rangle  &= \bigg( 0, \frac{1}{\sqrt{3}}Pk_+, -\sqrt{\frac{2}{3}}Pk_z, -Pk_-, \sqrt{\frac{2}{3}}Pk_+, -\frac{1}{\sqrt{3}}Pk_z \bigg) =  s^-_j
\end{align}

where $P$ is a matrix element computed on orbital states, defined in the Supplementary Information \ref{sec:effmass}, $k_z$ and $k_{\pm} = (k_x \pm ik_y )/\sqrt{2}$ are the electron wavevectors. When strain is introduced, the valence states mix and the Kane Hamiltonian must be rewritten in terms of the new valence states, which will be a superposition of the old ones $|l,m\rangle$. The new states can be found by diagonalizing the Orbit-Strain Hamiltonian  \cite{bir_symmetry_1974} which is

\begin{equation}
\label{eq:strainham}
            H_{\epsilon} = -a_c (\epsilon_{xx} + \epsilon_{yy} + \epsilon_{zz}) 
    - 3b_c \left[ \left( L_x^2 - \frac{L^2}{3} \right) \epsilon_{xx} + \text{c.p.} \right] 
    - \sqrt{3}d_c \left[ \left( (L_x L_y + L_y L_x) \epsilon_{xy} \right) + \text{c.p.} \right]
\end{equation}

where $L_i$ is the angular momentum operator, defined along the conventionally oriented reference frame, $b_c,d_c$ are the uniaxial deformation potential, $a_c$ is the deformation potential and c.p. stands for cyclic permutations over the $x,y,z$ indexes. We underline that in the calculation $\hbar$ is omitted since it is an overall multiplying term and is not required for the mass calculation. The full matrix form of the Hamiltonian in reported in Supplementary Information \ref{sec:effmass} but, already looking at eq.\ref{eq:strainham} it is clear that the calculation becomes non-trivial if the strain tensor presents non-diagonal terms, i.e. for any growth direction besides the standard (100). Indeed, for the latter case, Sugawara reports a closed analytical form of the new valence states \cite{sugawara_conduction-band_1993} that are a superposition of the states $\left|\frac{3}{2}, \pm \frac{1}{2}\right\rangle$ and $\left|\frac{1}{2}, \pm \frac{1}{2}\right\rangle$, where the mixing term effectively quantifies the strain in the system. Instead, for a generic growth direction, the strain tensor is not diagonal, therefore the system must be solved numerically. In general, it is possible to find that the new 6 states will still be degenerate in pairs, since the Orbit-Strain Hamiltonian does not act on the spin variable, and its eigenvectors can be written in terms of coefficients of the old base using the ordering of eq.\ref{eq:vastates}, as 

\begin{equation}
    v_i = (\alpha^i_1, \alpha_2^i, ...,\alpha^i_6) = \alpha^i_j
\end{equation}

where $i$ refers to the eigenvector number and $j$ to the coefficient number. Using this notation, and the  previously defined vectors $s^{\pm}_j$, it is possible to build the new Kane Hamiltonian, which will have the new 12 off-diagonal terms coupling the conduction states to the new valence states. The matrix elements can be expressed as $p_i^{\pm} = \alpha^i_js^{\pm}_j$ where the repeated indexes convention is used. Since we are anyway interested in the conduction band states, the 8x8 matrix can be folded into a 2x2 matrix using first order perturbation theory. In index notation, the matrix takes the form of:

\begin{equation}
    H^{2x2}_{ij} = \frac{^*p^{+}_kp^+_k}{E_g + P_\epsilon-E_k}\delta_{ij} + \frac{^*p^{-}_kp^+_k}{E_g + P_\epsilon-E_k}(1-\delta_{ij}) 
\end{equation}

where $E_k$ are the energies we retrieved from the Orbit-Strain Hamiltonian, $E_g$ is the band-gap and $P_\epsilon = \delta E_c$ and is kept in that notation for consistency with the work in \cite{sugawara_conduction-band_1993}. Upon carrying out the calculation, all off-diagonal terms vanish, independently of the specific form of the strain tensor, since the two $s_j^{\pm}$ vectors are orthogonal. The remaining step is then to write the diagonal term as a polynomial in $k_z, k_x, k_y$ to extract the electron mass. In the (100) configuration, this can be done straightforwardly, since the resulting polynomial is homogeneous and can thus be written in the form

\begin{equation}
    H^{2x2}_{ii} =  (a_1k_z^2+a_2k_x^2+a_3k_y^2)P^2
\end{equation}

Thus, the electron effective mass in the growth direction is

\begin{equation}
    \frac{m_0}{m_{eff}} = (1+D') + 2P^2a_1 \quad
    \text{where} \quad
    P^2 = \frac{1}{2} \left[ \frac{m_0}{m_{\Gamma}} - (1+D')\right] \frac{E_g(E_g + \Delta)}{E_g + \frac{2}{3}\Delta}
\end{equation}

where $D'=6$ is a second order correction term that is experimentally determined \cite{sugawara_conduction-band_1993}, $m_{\Gamma}$ is the effective mass at the $\Gamma$ point, $\Delta$ is the Split-Off energy and $m_0$ is free electron mass. Because all the remaining terms are independent of strain and orientation, the only quantity to be determined is the coefficient in front of $k_z^2$ in the final Hamiltonian. For the (511) orientation, the resulting polynomial is not homogeneous since the crystal axes are misaligned with the chosen coordinate frame. Consequently, we must rotate the coordinate system to retrieve the correct electron mass. A distinctive feature in this case is that the rotation does not yield an orthonormal basis aligned with the growth direction, but is tilted by approximately $\sim6^\circ$. This occurs because, as shown in the previous section, the strain tensor does not have eigenvectors that coincide with the growth direction. Therefore, at least in principle, when solving the Schrödinger equation for a heterostructure grown along such a direction, it is no longer possible to separate the in-plane transport from the transport in the out-of-plane direction. However, in practice, the difference in effective mass in the 3 main directions is not significantly large, in particular, for In$_{0.674}$Ga$_{0.326}$As we find $m_{eff}(511) = (0.0362, 0.0363, 0.0397)$, while for the (100) case we find $m_{eff}(100) = (0.0358, 0.0358, 0.0393)$. Specifically, looking at the $z$ component, the change is of about $\sim1\%$ which does not explain the measured red-shift, where we would have needed a change of about 50$\%$. The same procedure can be applied on the electron mass of Al$_{0.652}$In$_{0.348}$As, which shifts roughly by the same amount, but has a lower impact on the energy of the lasing transition. 

\section{Conclusions}

We have presented results on the growth of strain-compensated InGaAs/AlInAs QCLs on a substrate oriented along the (511)A direction by comparing AFM and XRD measurements. In general, we find that achieving a high material quality is more challenging in such direction compared to the conventional (100) substrate and it is not possible to grow good quality material on a (411)A substrate. We attribute these problems to the extremely narrow growth-parameter window and to the necessity of modulating the As flux depending on whether the well or the barrier material is grown because of the different incorporation on the growth surface. Nevertheless, it was still possible to obtain lasing devices, which we compare to the ones fabricated on the (100) material. Unfortunately, because of the different incorporation of Si along the (111) direction, which tends to increase the effective p-type background, the (511) laser has a lower doping. Even after HR coating, all the laser parameters result systematically lower than we the reference device. We also performed a spectral characterization which shows a 7\% red-shift of the emission wavelength. We attribute the redshift to an increase in impurity scattering, which is caused by the easier incorporation occurring along the exposed (111) direction on the (511) sample. In order to validate this statement, we perform a calculation on the change of CBO and electron effective mass for strained material along the (511) direction, giving a general procedure to evaluate these parameters for an arbitrary growth direction. We first prove that the change in the CBO is of the order of 1.2\% which cannot explain the 7\% wavelength shift. Then we compute the effective mass on for the (511) material finding that the electron transport does not align with the growth direction, presenting an angle of about $6^{\circ}$. In practice, this should not impact the usual transport approximation, since the change in electron mass remains around 1\%, when a change of about 50\% would be needed to match the change in wavelength. In conclusion, we believe that growing on non-trivial direction can still be a valid method to try to improve QCL performance by interface roughness control. However, looking at the data presented in this work, it is clear that the growth on strained material on such directions is more than challenging and does not bring, at least at the moment, to a significant performance increase. 

\newpage


\section*{Author contributions}
M. Beck calibrated, grew ancd characterized with XRD and AFM measurements the QCL heterostructures. A. Cargioli processed and characterized the lasers, and developed the calculations for the CBO and electron effective mass for strained material along non-trivial directions. J. Faist conceptualized and supervised the project. All authors contributed to the review of the manuscript.

\section*{Competing interests}
The authors declare no competing interest.

\section*{Supplementary Information}
Supplementary information is available 

\section*{Data Availability}
The data is available from the authors upon reasonable request.

\newpage

\section*{Epitaxial growth optimization, measurement and theoretical analysis of strain-compensated QCL grown on (511)A InP (Supplementary Material)}
\renewcommand{\figurename}{S}

\counterwithout{figure}{section}
\counterwithout{equation}{section}
\setcounter{figure}{0}

\section*{Scattering Rate calculation}
\label{supp:scattering}

To compute the interface roughness scattering rate between the upper (u) and lower (l) laser state, we use the expression reported in \cite{semtsiv_reduced_2018}

\begin{equation}
    R_{u,l} = \frac{\pi m_c}{\hbar^3} \Delta^2 \Lambda^2 \delta U^2 \eta \exp(-\Lambda^2 q^2 / 4)
\end{equation}

where $m_c = 0.04m_0$ is the effective electron mass, $\delta U = 827$meV is the conduction band offset, $\eta = 0.005$ nm$^{-2}$ is the product of electron probabilities summed over all interfaces, and $q$ is the scattering wavevector. We take these typical values, assuming a lasing transition of 270 meV ($\sim4.6\mu$m) in order to compare the difference between the scattering contribution for the (100) and (511)A case. Due to the (511) surface morphology we use a $\Delta = 0.22$nm and $\Lambda = 2.15$nm, while for the (100) case we use $\Delta = 0.29$ nm. For the latter, the correlation term is fixed by the growth condition, and it is hard to know a priori. Typical values can go from 4 to 12 nm, therefore strongly influencing the scattering rate at this emission wavelength.  

\newpage
\section*{Active Region Details}
\label{sec:active_region}

Strain compensated active region structure grown on both (100) and (511)A InP substrate. 

\begin{table}[htb!]
    \centering
    \renewcommand{\arraystretch}{1.2} 
     \resizebox{.5\textwidth}{!}{%
    \begin{tabular}{ccc}
        \toprule
        \multicolumn{3}{c}{\textbf{EV3032}} \\
        \cmidrule(lr){1-3} 
        
        Material & Thickness & Doping   \\
         & [\AA] & [$10^{17} \text{cm}^{-3}$]   \\
        \midrule
        
        \rowcolor{lightgray} 
        In$_{0.674}$Ga$_{0.326}$As & 27 &  \\
        
        Al$_{0.652}$In$_{0.348}$As & 19 &   \\
        
        \rowcolor{lightgray} 
        InGaAs & 26 & \\
        
        AlInAs & 15 &  \\
        
        \rowcolor{lightgray} 
        InGaAs & 23 &   \\
        
        AlInAs & 14 &  \\
        
        \rowcolor{lightgray} 
        InGaAs & 21 & 1.53 \\
        
        AlInAs & 22 & 0.823  \\
        
        \rowcolor{lightgray} 
        InGaAs & 19 & 1.53 \\
        
        AlInAs & 20 & 0.823 \\
        
        \rowcolor{lightgray} 
        InGaAs & 19 & 1.53 \\
        
        AlInAs & 19 &  \\
        
        \rowcolor{lightgray} 
        InGaAs & 17 &\\
        
        AlInAs & 24 &\\
        
        \rowcolor{lightgray} 
        InGaAs & 17 & \\
        
        AlInAs & 35 & \\

         \rowcolor{lightgray} 
        InGaAs & 11 & \\
        
        AlInAs & 13 &  \\

         \rowcolor{lightgray} 
        InGaAs & 38 & \\
        
        AlInAs & 10 & \\

         \rowcolor{lightgray} 
        InGaAs & 35 & \\
        
        AlInAs & 18 &  \\

        \bottomrule
    \end{tabular}%
    }
\end{table}

\section*{AFM Measurements on Bulk Mayterial}
\label{sec:afm}

\begin{figure}[htb!]
    \centering
    \includegraphics[width=.85\linewidth]{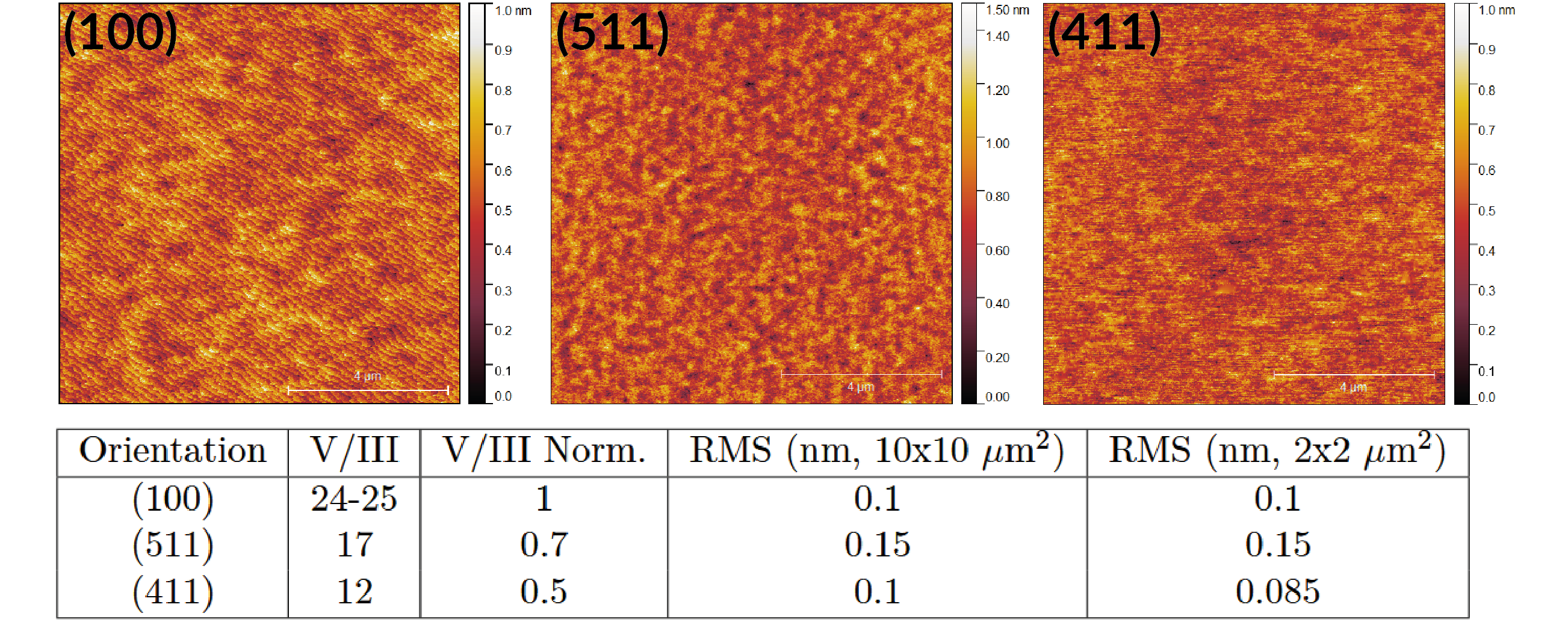}
    \caption{AFM measurements of the surface roughness of 0.5$\mu$m bulk InGaAs grown on (100), (511) and (411) crystal direction. V/III flux ratios are reported as bare fluxes and normalized to the (100). RMS values of the surface roughness are also shown for 2 different integration areas.}
    \label{fig:bulkInGaAs}
\end{figure}

\begin{figure}[htb!]
    \centering
    \includegraphics[width=.85\linewidth]{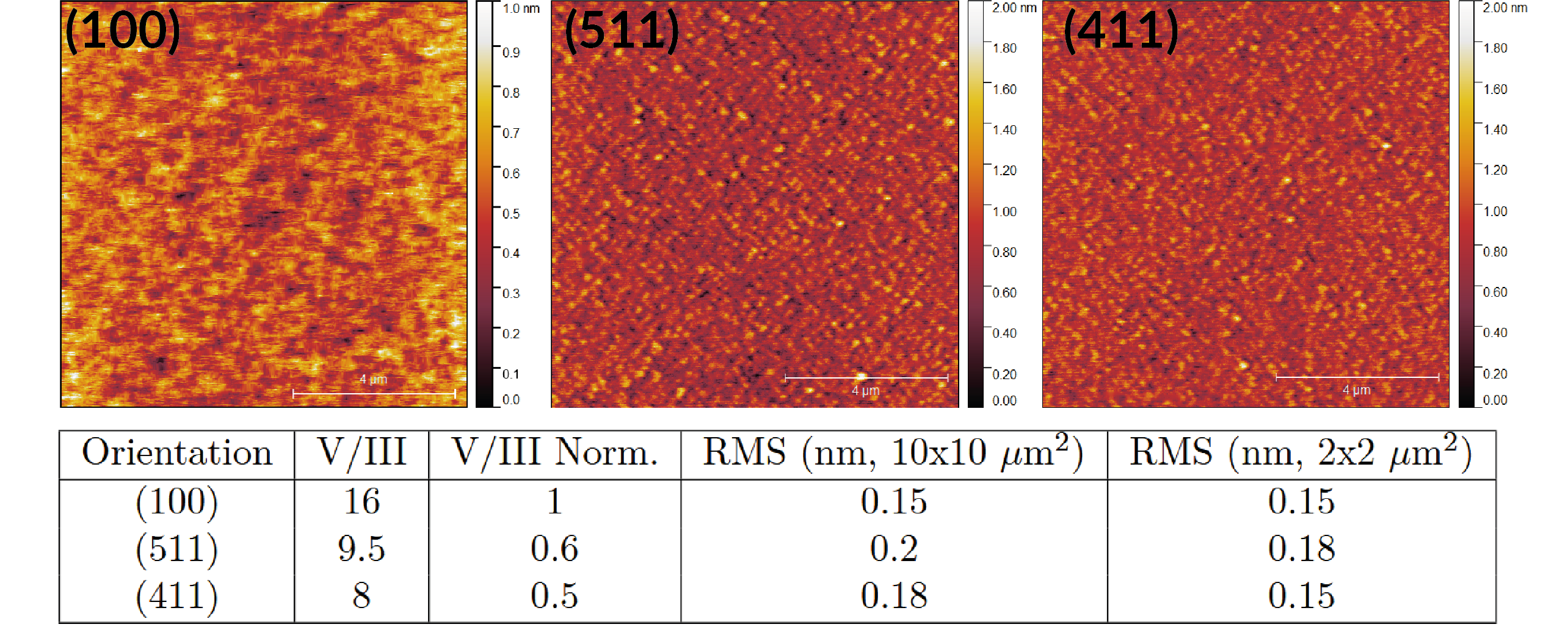}
    \caption{AFM measurements of the surface roughness of 0.5$\mu$m bulk AlInAs grown on (100), (511) and (411) crystal direction. V/III flux ratios are reported as bare fluxes and normalized to the (100). RMS values of the surface roughness are also shown for 2 different integration areas.}
    \label{fig:bulkAlInAs}
\end{figure}

In Figure S\ref{fig:bulkInGaAs}-\ref{fig:bulkAlInAs}, we report the AFM measurements for 0.5$\mu$m thick InGaAs and AlInAs, respectively. Because of the terrace morphology, the (411)A and (511)A orientations tend to incorporate a higher amount of As, so the V/III ratio must be carefully adjusted to obtain the desired material composition. As shown in the corresponding tables, the surface roughness of InGaAs in the (411) direction appears comparable to, or even better than, that of the (100) orientation, whereas for the (511)A direction the roughness is slightly worse. Nevertheless, it was not possible to obtain strain-compensated material on the (411)A substrate without relaxation. Therefore QCLs were grown only on the (511)A substrate. 

\section*{Laser Processing}
\label{sec:laser proc}

The lasers are processed using a standard ridge configuration. Specifically, the waveguides are wet-etched using an HBr:Br:H$_2$O solution and passivated with SiNx. The top of the waveguide is then opened using Reactive Ion Etching and contacted via Ti/Pt/Au (10/40/150 nm). Subsequently, a 3 to 4 $\mu$m thick gold layer is electroplated on top of the lasers to ensure easier bonding and better heat dissipation. We emphasize that because of the unusual growth orientation, cleaving lasers fabricated from the (511)A sample is not straightforward. Specifically, the waveguide must be aligned parallel to the big-flat, i.e., along the ($\bar{1}$10) direction. This is necessary because cleaving perpendicular to the big-flat still yields 90$^{\circ}$ facets, while cleaving parallel to it results in an angle of approximately 19-20 $^{\circ}$, which would lower the facet reflectivity. Moreover, it was observed that cleaving along this direction requires approximately 2 to 3 times more force than for a standard substrate.

\section*{Strain Tensor}
\label{sec:cbo}

The $A$ to $F$ parameters in the tensor espression can be written as 

\begin{align}
A &= c_{44}(c_{11} + 2c_{12})(-c_{11} + c_{12} + c_{44}), \\
B &= c_{44}(c_{11} + 2c_{12})(c_{11} - c_{12} - 2c_{44}), \\
C &= -(c_{11} + 2c_{12})(c_{11} - c_{12} - 2c_{44})^2, \\
D &= \frac{1}{4}c_{44}(c_{11}^2 - c_{12}^2 - 2c_{12}c_{44}), \\
E &= -\frac{1}{2}c_{44}(c_{11} + c_{12})(c_{11} - c_{12} - 2c_{44}), \\
F &= (c_{11} - c_{12} - 2c_{44})^2(c_{11} + 2c_{12} + c_{44}).
\end{align}

where $c_{11},c_{12},c_{44} $ are the 3 independent elastic constant used to describe cubic materials. The full strain tensor in the standard cartesian base is given by  \\

\begin{equation}
\resizebox{\textwidth}{!}{$
  \begin{pmatrix}
    \frac{4 c_{44} (175 c_{11} - 13 c_{12} - c_{44})}{25 c_{11}^3 - 25 c_{11}^2 c_{12} - 58 c_{11}^2 c_{44} - 677 c_{11} c_{44} - 98 c_{12} c_{44} - 4 c_{44}^2} & -\frac{(c_{11} + 2 c_{12}) (25 c_{11} - 25 c_{12} - 23 c_{44})}{25 c_{11}^3 - 25 c_{11}^2 c_{12} - 58 c_{11}^2 c_{44} - 677 c_{11} c_{44} - 98 c_{12} c_{44} - 4 c_{44}^2} & \frac{5 (c_{11} + 2 c_{12}) (13 c_{11} - 13 c_{12} - c_{44})}{25 c_{11}^3 - 25 c_{11}^2 c_{12} - 58 c_{11}^2 c_{44} - 677 c_{11} c_{44} - 98 c_{12} c_{44} - 4 c_{44}^2} \\
    -\frac{(c_{11} + 2 c_{12}) (25 c_{11} - 25 c_{12} - 23 c_{44})}{25 c_{11}^3 - 25 c_{11}^2 c_{12} - 58 c_{11}^2 c_{44} - 677 c_{11} c_{44} - 98 c_{12} c_{44} - 4 c_{44}^2} & \frac{4 c_{44} (175 c_{11} - 13 c_{12} - c_{44})}{25 c_{11}^3 - 25 c_{11}^2 c_{12} - 58 c_{11}^2 c_{44} - 677 c_{11} c_{44} - 98 c_{12} c_{44} - 4 c_{44}^2} & -\frac{5 (c_{11} + 2 c_{12}) (13 c_{11} - 13 c_{12} - c_{44})}{25 c_{11}^3 - 25 c_{11}^2 c_{12} - 58 c_{11}^2 c_{44} - 677 c_{11} c_{44} - 98 c_{12} c_{44} - 4 c_{44}^2} \\
    -\frac{5 (c_{11} + 2 c_{12}) (13 c_{11} - 13 c_{12} - c_{44})}{25 c_{11}^3 - 25 c_{11}^2 c_{12} - 58 c_{11}^2 c_{44} - 677 c_{11} c_{44} - 98 c_{12} c_{44} - 4 c_{44}^2} & -\frac{5 (c_{11} + 2 c_{12}) (13 c_{11} - 13 c_{12} - c_{44})}{25 c_{11}^3 - 25 c_{11}^2 c_{12} - 58 c_{11}^2 c_{44} - 677 c_{11} c_{44} - 98 c_{12} c_{44} - 4 c_{44}^2} & \frac{4 c_{44} (13 c_{11} - 337 c_{12} - c_{44})}{25 c_{11}^3 - 25 c_{11}^2 c_{12} - 58 c_{11}^2 c_{44} - 677 c_{11} c_{44} - 98 c_{12} c_{44} - 4 c_{44}^2}
  \end{pmatrix}$
}
\end{equation}

The diagonalization of the matrix provides two non-trivial eigenvalues in the form $\lambda_1\pm\lambda_2$. We define here the two values as functions of the elastic constants.

\begin{align}
   \lambda_1 &=  \frac{25 c_{11}^2 - 58 c_{12}^2 - 25 c_{11} (c_{12} - 31 c_{44}) - 1354 c_{12} c_{44} - 8 c_{44}^2}{2 \left( 25 c_{11}^3 - 25 c_{11}^2 c_{12} - 58 c_{11}^2 c_{44} - 677 c_{11} c_{44} - 98 c_{12} c_{44} - 4 c_{44}^2 \right)} \\
   \lambda_2 &= \frac{9 \sqrt{(c_{11} - 2 c_{12})^2 (425 c_{11}^2 - 425 c_{12}^2 - 358 c_{12} c_{44} - 5561 c_{44}^2 - 58 c_{11} (17 c_{12} - 7 c_{44}))}}{2 \left( 25 c_{11}^3 - 25 c_{11}^2 c_{12} - 58 c_{11}^2 c_{44} - 677 c_{11} c_{44} - 98 c_{12} c_{44} - 4 c_{44}^2 \right)}
\end{align}

The two non-trivial eigenvectors are also written with components in the form of $\Lambda_1\pm\Lambda_2$ where

\begin{align}
    \Lambda_1 &= \frac{(c_{11} + 2 c_{12}) (25 c_{11} - 25 c_{12} - 671 c_{44})}{20 (c_{11} + 2 c_{12}) (13 c_{11} - 13 c_{12} + c_{44})}\\
    \Lambda_2 &= \frac{9 \sqrt{(c_{11} + 2 c_{12})^2 (425 (c_{11} - c_{12})^2 + 358 (-c_{11} + c_{12}) c_{44} + 5561 c_{44}^2)}}{20 (c_{11} + 2 c_{12}) (13 c_{11} - 13 c_{12} + c_{44})}
\end{align}

The full form for the trace of the tensor is 

\begin{equation}
    Tr(\epsilon_{ij}) = \frac{12 c_{44} (121 c_{11} - 121 c_{12} + c_{44})}{25 c_{11}^2 + 25 c_{11} c_{12} - 50 c_{12}^2 + 677 c_{11} c_{44} - 98 c_{12} c_{44} + 4 c_{44}^2}
\end{equation}

\subsection{Material Parameters}
\label{ap:matpar}

All parameters are calculated using the virtual crystal approach, starting from the data of bulk InAs, AlAs, and GaAs tabulated in \cite{van_de_walle_band_1989, vurgaftman_band_2001}.

\begin{table}[h!]
\centering
\begin{tabular}{l|c|c}
Parameter & Al$_{0.652}$In$_{0.348}$As & In$_{0.674}$Ga$_{0.326}$As \\
\hline
$c_{11}$ & 1.1049 & 0.93898 \\
$c_{12}$ & 0.50581 & 0.48132 \\
$c_{44}$ & 0.49119 & 0.46087 \\
$m_0/m_{\Gamma}$ & 0.087044 & 0.032618 \\
$E_g$ & 1.9223 & 0.59738 \\
$\Delta$ & 0.28425 & 0.34107 \\
$a_c$ & -6.125 & -6.3348 \\
$b_c$ & -0.6264 & -1.7674 \\
$d_c$ &  -3.4696 & -3.8564 \\
$\epsilon_\parallel $& 5.8687/5.7993 - 1 & 5.8687 / 5.9262537 - 1 \\
\end{tabular}
\caption{Parameters used to perform the CBO and effective mass calculation.}
\label{tab:material_parameters}
\end{table}

\section*{Effective Mass Calculation}
\label{sec:effmass}

For the conduction band, we define $|S\uparrow\rangle$ and $|S\downarrow\rangle$ as the two conduction band states, where the arrows indicate the spin state and $S$ the orbital part of the wavefunction. Instead, for the valence band, the states are chosen as

\begin{equation}
\begin{array}{rcl@{\hspace{2em}}rcl}
\left|\frac{3}{2}, \frac{3}{2}\right\rangle & = & \frac{1}{\sqrt{2}}|(X+iY)\uparrow\rangle & \left|\frac{3}{2}, -\frac{3}{2}\right\rangle & = & -\frac{1}{\sqrt{2}}|(X-iY)\downarrow\rangle \\
\left|\frac{3}{2}, \frac{1}{2}\right\rangle & = & \frac{1}{\sqrt{6}}|(X+iY)\downarrow - 2Z\uparrow\rangle & \left|\frac{3}{2}, -\frac{1}{2}\right\rangle & = & -\frac{1}{\sqrt{6}}|(X-iY)\uparrow + 2Z\downarrow\rangle \\
\left|\frac{1}{2}, \frac{1}{2}\right\rangle & = & \frac{1}{\sqrt{3}}|(X+iY)\downarrow + Z\uparrow\rangle & \left|\frac{1}{2}, -\frac{1}{2}\right\rangle & = & \frac{1}{\sqrt{3}}|(X-iY)\uparrow - Z\downarrow\rangle
\end{array}
\end{equation}

where $X,Y,Z$ are $p$ orbitals. The $\left|\frac{3}{2}, \pm \frac{3}{2}\right\rangle$ states are associated with the Heavy-Hole band, $\left|\frac{3}{2}, \pm \frac{1}{2}\right\rangle$ with the Light-Hole band, and  $\left|\frac{1}{2}, \pm \frac{1}{2}\right\rangle$ with the Split-Off band. The full matrix form of the Kane Hamiltonian is then computed by evaluating all matrix elements on the $\textbf{k}\cdot\textbf{p}$ operator, 

\begin{equation}
\label{eq:Kane}
    \begin{pmatrix}
    S^+  & HH^+ & LH^+ & SO^+ & S^- &HH^- & LH^- & SO^-\\
    0 & -\sqrt{\frac{2}{3}} P  k_z & P  k_+ & \frac{1}{\sqrt{3}} P  k_z & 0 & -\frac{1}{\sqrt{3}} P  k_- & 0 & -\sqrt{\frac{2}{3}} P  k_- \\
    -\sqrt{\frac{2}{3}}P  k_z & -E_g & 0 & 0 & \frac{1}{\sqrt{3}} P  k_- & 0 & 0 & 0 \\
    P  k_- & 0 & -E_g & 0 & 0 & 0 & 0 & 0 \\
    \frac{1}{\sqrt{3}} P  k_z & 0 & 0 & -E_g - \Delta & \sqrt{\frac{2}{3}} P  k_- & 0 & 0 & 0 \\
    0 & \frac{1}{\sqrt{3}} P  k_+ & 0 & \sqrt{\frac{2}{3}} P  k_+ & 0 & -\sqrt{\frac{2}{3}} P  k_z & P  k_- & \frac{1}{\sqrt{3}} P  k_z \\
    -\frac{1}{\sqrt{3}} P  k_+ & 0 & 0 & 0 & -\sqrt{\frac{2}{3}} P  k_z & -E_g & 0 & 0 \\
    0 & 0 & 0 & 0 & P  k_+ & 0 & -E_g & 0 \\
    -\sqrt{\frac{2}{3}} P  k_+ & 0 & 0 & \frac{1}{\sqrt{3}} P  k_z & 0 & 0 & 0 & -E_g - \Delta
    \end{pmatrix}
\end{equation}

where the 0 of the energy is chosen at the bottom of the conduction band, $E_g$ is the band-gap, $\Delta$ is the Split-Off energy and the rest of the parameters are defined as

\begin{equation}
    k_{\pm} = \frac{1}{\sqrt{2}}(k_x \pm i k_y) \quad P = -\frac{i}{m_0} \langle S|p_x|X \rangle = -\frac{i}{m_0} \langle S|p_y|Y \rangle = -\frac{i}{m_0} \langle S|p_z|Z \rangle.
\end{equation}

The ordering of the Hamiltonian is reported in the first line of the matrix and follows the convention used in \cite{sugawara_conduction-band_1993}.

 The strain Hamiltonian over the 6 valence states is written as

\begin{equation}
\label{eq:orbitstrain}
H_{\epsilon} = -
\left(
\begin{array}{cccccc}
\textstyle \left|\frac{3}{2}, \frac{3}{2}\right\rangle & \textstyle \left|\frac{3}{2}, \frac{1}{2}\right\rangle & \textstyle \left|\frac{3}{2}, -\frac{1}{2}\right\rangle & \textstyle \left|\frac{3}{2}, -\frac{3}{2}\right\rangle & \textstyle \left|\frac{1}{2}, \frac{1}{2}\right\rangle & \textstyle \left|\frac{1}{2}, -\frac{1}{2}\right\rangle \\[1em]
P_{\epsilon} + Q_{\epsilon} & -S_{\epsilon} & R_{\epsilon} & 0 & \frac{1}{\sqrt{2}} S_{\epsilon} & -\sqrt{2} R_{\epsilon} \\
-S_{\epsilon}^{\ast} & P_{\epsilon} - Q_{\epsilon} & 0 & R_{\epsilon} & \sqrt{2} Q_{\epsilon} & -\sqrt{\frac{3}{2}} S_{\epsilon} \\
R_{\epsilon}^{\ast} & 0 & P_{\epsilon} - Q_{\epsilon} & S_{\epsilon} & -\sqrt{\frac{3}{2}} S_{\epsilon}^{\ast} & -\sqrt{2} Q_{\epsilon} \\
0 & R_{\epsilon}^{\ast} & S_{\epsilon}^{\ast} & P_{\epsilon} + Q_{\epsilon} & \sqrt{2} R_{\epsilon}^{\ast} & \frac{1}{\sqrt{2}} S_{\epsilon}^{\ast} \\
\frac{1}{\sqrt{2}} S_{\epsilon}^{\ast} & \sqrt{2} Q_{\epsilon} & -\sqrt{\frac{3}{2}} S_{\epsilon} & \sqrt{2} R_{\epsilon} & P_{\epsilon} + \Delta & 0 \\
-\sqrt{2} R_{\epsilon}^{\ast} & -\sqrt{\frac{3}{2}} S_{\epsilon}^{\ast} & -\sqrt{2} Q_{\epsilon} & \frac{1}{\sqrt{2}} S_{\epsilon} & 0 & P_{\epsilon} + \Delta
\end{array}
\right)
\end{equation}

where the elements of the matrix are functions of the strain tensor components

\begin{align}
P_{\epsilon} &= a_c (\epsilon_{xx} + \epsilon_{yy} + \epsilon_{zz}) \\
Q_{\epsilon} &= \frac{b_c}{2} (2\epsilon_{zz} - \epsilon_{xx} - \epsilon_{yy}) \\
S_{\epsilon} &= -(\epsilon_{zx} - i\epsilon_{yz})d_c \\
R_{\epsilon} &= \frac{\sqrt{3}}{2} b_c (\epsilon_{xx} - \epsilon_{yy}) - i d_c \epsilon_{xy}
\end{align}

where $b_c,d_c$ are the uniaxial deformation potential, $a_c$ is the hydrostatic deformation potential. The strain tensor components are defined by eq.\ref{eq:straintensor} of the main text. We intentionally omit the $\hbar$ term for readability since it is a constant multiplying the full matrix and it is not important for calculating the effective mass. After applying the diagonalization procedure described in the main text, it is possible to retrieve the polynomials in $k_{x,y,z}$ and extract the electron mass. For completeness we report the numerical results obtained by this procedure. For the (100) case the polynomial is already homogeneous and the coefficients are 

\begin{equation}
    H^{2x2}_{ii}(100)= P^2(1.14662k_z^2 + 1.29992 k_x^2 + 1.29992  k_y^2)
\end{equation}

Instead, for the (511) case we find, in the same coordinate system:

\begin{align}
    H^{2x2}_{ii}(511)= P^2(1.13857 k_z^2 - &0.0338399 k_zk_x + \\
    1.28154 k_x^2 -& 0.0338399 k_zk_y - 0.0006856 k_x k_y + 1.28154 k_y^2)
\end{align}

Defining as $(k'_x,k'_y,k'_z)$ the wavevectors in the reference frame where the polynomial becomes homogeneous we find

\begin{align}
    &H^{2x2}_{ii}(511)= P^2( 1.13466 k^{'2}_z + 1.2851 k^{'2}_x +1.28188 k^{'2}_y \\
    &k'_z = (0.113955, 0.113955, 0.986929) \\
    &k'_y = ({0.707107, -0.707107, 0}) \\
    &k'_x = (-0.697864, -0.697864, 0.161156) 
\end{align}

where we expressed the new $k'$ basis in terms of the old one. We underline that the set of eigenvectors $k'$ is the same set that we would find by diagonalizing the stress tensor with the parameters of (511), and that the scalar product of $k'_z$ with a unit vector oriented along the (511) allows us to estimate an angle of about $\sim6^\circ$.

\bibliographystyle{ieeetr}
\bibliography{references}

@article{tsujino_interface-roughness-induced_2005,
	title = {Interface-roughness-induced broadening of intersubband electroluminescence in p-{SiGe} and n-{GaInAs}∕{AlInAs} quantum-cascade structures},
	volume = {86},
	issn = {0003-6951},
	url = {https://doi.org/10.1063/1.1862344},
	doi = {10.1063/1.1862344},
	number = {6},
	urldate = {2026-01-30},
	journal = {Applied Physics Letters},
	author = {Tsujino, S. and Borak, A. and Müller, E. and Scheinert, M. and Falub, C. V. and Sigg, H. and Grützmacher, D. and Giovannini, M. and Faist, J.},
	month = feb,
	year = {2005},
	pages = {062113},
}

@article{sugawara_conduction-band_1993,
	title = {Conduction-band and valence-band structures in strained {In}\{1-x\}{Ga}(x){As}/{InP} quantum wells on (001) {InP} substrates},
	volume = {48},
	url = {https://link.aps.org/doi/10.1103/PhysRevB.48.8102},
	doi = {10.1103/PhysRevB.48.8102},
	number = {11},
	urldate = {2025-09-30},
	journal = {Physical Review B},
	publisher = {American Physical Society},
	author = {Sugawara, Mitsuru and Okazaki, Niroh and Fujii, Takuya and Yamazaki, Susumu},
	month = sep,
	year = {1993},
	pages = {8102--8118},
}

@book{bir_symmetry_1974,
	title = {Symmetry and {Strain}-induced {Effects} in {Semiconductors}},
	isbn = {978-0-470-07321-6},
	language = {en},
	publisher = {Wiley},
	author = {Bir, Gennadiĭ Levikovich and Pikus, Grigoriĭ Ezekielevich},
	year = {1974},
	note = {Google-Books-ID: 38m2QgAACAAJ},
}

@article{kane_band_1957,
	title = {Band structure of indium antimonide},
	volume = {1},
	issn = {0022-3697},
	url = {https://www.sciencedirect.com/science/article/pii/0022369757900136},
	doi = {10.1016/0022-3697(57)90013-6},
	number = {4},
	urldate = {2025-11-07},
	journal = {Journal of Physics and Chemistry of Solids},
	author = {Kane, Evan O.},
	month = jan,
	year = {1957},
	pages = {249--261},
}

@article{kane_energy_1956,
	title = {Energy band structure in p-type germanium and silicon},
	volume = {1},
	issn = {0022-3697},
	url = {https://www.sciencedirect.com/science/article/pii/0022369756900142},
	doi = {10.1016/0022-3697(56)90014-2},
	number = {1},
	urldate = {2025-11-07},
	journal = {Journal of Physics and Chemistry of Solids},
	author = {Kane, E. O.},
	month = sep,
	year = {1956},
	pages = {82--99},
}

@article{masataka_higashiwaki_dc_2000,
	title = {{DC} and {RF} {Performance} of 50 nm {Gate} {Pseudomorphic} {In}$_{\textrm{0.7}}$ {Ga}$_{\textrm{0.3}}$ {As}/{In}$_{\textrm{0.52}}$ {Al}$_{\textrm{0.48}}$ {As} {High} {Electron} {Mobility} {Transistors} {Grown} on (411){A}-{Oriented} {InP} {Substrates} by {Molecular}-{Beam} {Epitaxy}},
	volume = {39},
	issn = {0021-4922, 1347-4065},
	url = {https://iopscience.iop.org/article/10.1143/JJAP.39.L720},
	doi = {10.1143/JJAP.39.L720},
	language = {en},
	number = {7B},
	urldate = {2025-11-05},
	journal = {Japanese Journal of Applied Physics},
	author = {Masataka Higashiwaki, Masataka Higashiwaki and Takahiro Kitada, Takahiro Kitada and Toyohiro Aoki, Toyohiro Aoki and Satoshi Shimomura, Satoshi Shimomura and Yoshimi Yamashita, Yoshimi Yamashita and Akira Endoh, Akira Endoh and Kohki Hikosaka, Kohki Hikosaka and Takashi Mimura, Takashi Mimura and Toshiaki Matsui, Toshiaki Matsui and Satoshi Hiyamizu, Satoshi Hiyamizu},
	month = jul,
	year = {2000},
	pages = {L720},
}

@article{shimomura_extremely_1995,
	title = {Extremely flat interfaces in {GaAs}/{AlGaAs} quantum wells with high {Al} content (0.7) grown on {GaAs} (411){A} substrates by molecular beam epitaxy},
	volume = {150},
	issn = {0022-0248},
	url = {https://www.sciencedirect.com/science/article/pii/0022024895802447},
	doi = {10.1016/0022-0248(95)80244-7},
	urldate = {2025-11-05},
	journal = {Journal of Crystal Growth},
	author = {Shimomura, Satoshi and Kaneko, Shinjiroh and Motokawa, Takeharu and Shinohara, Keisuke and Adachi, Akira and Okamoto, Yasunori and Sano, Naokatsu and Murase, Kazuo and Hiyamizu, Satoshi},
	month = may,
	year = {1995},
	pages = {409--414},
}

@article{hiyamizu_extremely_1994,
	title = {Extremely high uniformity of interfaces in {GaAs}/{AlGaAs} quantum wells grown on (411){A} {GaAs} substrates by molecular beam epitaxy},
	volume = {12},
	issn = {1071-1023, 1520-8567},
	url = {https://pubs.aip.org/jvb/article/12/2/1043/587171/Extremely-high-uniformity-of-interfaces-in-GaAs},
	doi = {10.1116/1.587082},
	language = {en},
	number = {2},
	urldate = {2025-11-05},
	journal = {Journal of Vacuum Science \& Technology B: Microelectronics and Nanometer Structures Processing, Measurement, and Phenomena},
	author = {Hiyamizu, S. and Shimomura, S. and Wakejima, A. and Kaneko, S. and Adachi, A. and Okamoto, Y. and Sano, N. and Murase, K.},
	month = mar,
	year = {1994},
	pages = {1043--1046},
}

@article{unuma_intersubband_2003,
	title = {Intersubband absorption linewidth in {GaAs} quantum wells due to scattering by interface roughness, phonons, alloy disorder, and impurities},
	volume = {93},
	language = {en},
	number = {3},
	journal = {J. Appl. Phys.},
	author = {Unuma, Takeya and Yoshita, Masahiro and Noda, Takeshi and Sakaki, Hiroyuki and Akiyama, Hidefumi},
	year = {2003},
}

@article{yang_strain_1994,
	title = {Strain in pseudomorphic films grown on arbitrarily oriented substrates},
	volume = {65},
	issn = {0003-6951, 1077-3118},
	url = {https://pubs.aip.org/apl/article/65/22/2789/63813/Strain-in-pseudomorphic-films-grown-on-arbitrarily},
	doi = {10.1063/1.112564},
	language = {en},
	number = {22},
	urldate = {2024-06-17},
	journal = {Applied Physics Letters},
	author = {Yang, Kai and Anan, Takayoshi and Schowalter, Leo J.},
	month = nov,
	year = {1994},
	pages = {2789--2791},
}

@article{vurgaftman_band_2001,
	title = {Band parameters for {III}–{V} compound semiconductors and their alloys},
	volume = {89},
	issn = {0021-8979, 1089-7550},
	url = {https://pubs.aip.org/jap/article/89/11/5815/488612/Band-parameters-for-III-V-compound-semiconductors},
	doi = {10.1063/1.1368156},
	language = {en},
	number = {11},
	urldate = {2024-04-24},
	journal = {Journal of Applied Physics},
	author = {Vurgaftman, I. and Meyer, J. R. and Ram-Mohan, L. R.},
	month = jun,
	year = {2001},
	pages = {5815--5875},
}

@article{van_de_walle_band_1989,
	title = {Band lineups and deformation potentials in the model-solid theory},
	volume = {39},
	copyright = {http://link.aps.org/licenses/aps-default-license},
	issn = {0163-1829},
	url = {https://link.aps.org/doi/10.1103/PhysRevB.39.1871},
	doi = {10.1103/PhysRevB.39.1871},
	language = {en},
	number = {3},
	urldate = {2024-04-24},
	journal = {Physical Review B},
	author = {Van De Walle, Chris G.},
	month = jan,
	year = {1989},
	pages = {1871--1883},
}

@article{khurgin_inhomogeneous_2008,
	title = {Inhomogeneous origin of the interface roughness broadening of intersubband transitions},
	volume = {93},
	issn = {0003-6951, 1077-3118},
	url = {https://pubs.aip.org/apl/article/93/9/091104/764971/Inhomogeneous-origin-of-the-interface-roughness},
	doi = {10.1063/1.2977994},
	language = {en},
	number = {9},
	urldate = {2023-11-27},
	journal = {Applied Physics Letters},
	author = {Khurgin, Jacob B.},
	month = sep,
	year = {2008},
	pages = {091104},
}

@article{semtsiv_reduced_2018,
	title = {Reduced interface roughness scattering in {InGaAs}/{InAlAs} quantum cascade lasers grown on (411){A} {InP} substrates},
	volume = {113},
	issn = {0003-6951, 1077-3118},
	url = {https://pubs.aip.org/apl/article/113/12/121110/986038/Reduced-interface-roughness-scattering-in-InGaAs},
	doi = {10.1063/1.5049090},
	language = {en},
	number = {12},
	urldate = {2023-10-06},
	journal = {Applied Physics Letters},
	author = {Semtsiv, M. P. and Kurlov, S. S. and Alcer, D. and Matsuoka, Y. and Kischkat, J.-F. and Bierwagen, O. and Masselink, W. T.},
	month = sep,
	year = {2018},
	pages = {121110},
}

@article{deutsch_probing_2013,
	title = {Probing scattering mechanisms with symmetric quantum cascade lasers},
	volume = {21},
	issn = {1094-4087},
	url = {https://opg.optica.org/oe/abstract.cfm?uri=oe-21-6-7209},
	doi = {10.1364/OE.21.007209},
	language = {en},
	number = {6},
	urldate = {2023-09-25},
	journal = {Optics Express},
	author = {Deutsch, Christoph and Detz, Hermann and Zederbauer, Tobias and Andrews, Aaron M. and Klang, Pavel and Kubis, Tillmann and Klimeck, Gerhard and Schuster, Manfred E. and Schrenk, Werner and Strasser, Gottfried and Unterrainer, Karl},
	month = mar,
	year = {2013},
	pages = {7209},
}

@article{franckie_impact_2015,
	title = {Impact of interface roughness distributions on the operation of quantum cascade lasers},
	volume = {23},
	issn = {1094-4087},
	url = {https://opg.optica.org/abstract.cfm?URI=oe-23-4-5201},
	doi = {10.1364/OE.23.005201},
	language = {en},
	number = {4},
	urldate = {2023-09-25},
	journal = {Optics Express},
	author = {Franckié, Martin and Winge, David O. and Wolf, Johanna and Liverini, Valeria and Dupont, Emmanuel and Trinité, Virginie and Faist, Jérôme and Wacker, Andreas},
	month = feb,
	year = {2015},
	pages = {5201},
}

@article{flores_leakage_2013,
	title = {Leakage current in quantum-cascade lasers through interface roughness scattering},
	volume = {103},
	issn = {0003-6951, 1077-3118},
	url = {https://pubs.aip.org/apl/article/103/16/161102/25739/Leakage-current-in-quantum-cascade-lasers-through},
	doi = {10.1063/1.4825229},
	language = {en},
	number = {16},
	urldate = {2023-09-25},
	journal = {Applied Physics Letters},
	author = {Flores, Y. V. and Kurlov, S. S. and Elagin, M. and Semtsiv, M. P. and Masselink, W. T.},
	month = oct,
	year = {2013},
	pages = {161102},
}

@article{faist_quantum_1994,
	title = {Quantum {Cascade} {Laser}},
	volume = {264},
	url = {https://www.science.org/doi/abs/10.1126/science.264.5158.553},
	doi = {10.1126/science.264.5158.553},
	number = {5158},
	journal = {Science},
	author = {Faist, Jerome and Capasso, Federico and Sivco, Deborah L. and Sirtori, Carlo and Hutchinson, Albert L. and Cho, Alfred Y.},
	year = {1994},
	note = {\_eprint: https://www.science.org/doi/pdf/10.1126/science.264.5158.553},
	pages = {553--556},
}

@book{faist_quantum_2013,
	title = {Quantum {Cascade} {Lasers}},
	isbn = {978-0-19-852824-1},
	url = {https://academic.oup.com/book/3819},
	urldate = {2023-06-09},
	publisher = {Oxford University Press},
	author = {Faist, Jérôme},
	month = mar,
	year = {2013},
	doi = {10.1093/acprof:oso/9780198528241.001.0001},
	doi = {10.1093/acprof:oso/9780198528241.001.0001},
}

@article{bismuto_influence_2011,
	title = {Influence of the growth temperature on the performances of strain-balanced quantum cascade lasers},
	volume = {98},
	issn = {0003-6951, 1077-3118},
	url = {http://aip.scitation.org/doi/10.1063/1.3561754},
	doi = {10.1063/1.3561754},
	language = {en},
	number = {9},
	urldate = {2022-08-02},
	journal = {Applied Physics Letters},
	author = {Bismuto, A. and Terazzi, R. and Beck, M. and Faist, Jerome},
	month = feb,
	year = {2011},
	pages = {091105},
}

@article{khurgin_role_nodate,
	title = {Role of interface roughness in the transport and lasing characteristics of quantum-cascade lasers},
	language = {en},
	journal = {Appl. Phys. Lett.},
	author = {Khurgin, Jacob B and Dikmelik, Yamac and Liu, Peter Q},
	pages = {4},
}

\end{document}